\documentclass[doublecol]{epl2} 

\usepackage{lineno}
\usepackage{siunitx}
\usepackage{float}
\usepackage{lipsum}

\title{Beam focus modifications by cropping partially coherent X-ray beams}
\shorttitle{cropped X-ray beams} 

\author{Manuel Sanchez del Rio\inst{1} \and Rafael Celestre\inst{1} \and Juan Reyes-Herrera\inst{1} \and Philipp Brumund\inst{1} \and \newline Marco Cammarata\inst{1}}
\shortauthor{M. Sanchez del Rio \etal}

\institute{                    
  \inst{1} European Synchrotron Radiation Facility, 71 Avenue des Martyrs, 38000 Grenoble, France
}

\abstract{
We simulate the focusing of a partially-coherent X-ray beam emitted by an undulator in a fourth-generation storage ring by performing a coherent mode decomposition and wave optics propagation. The focus position is shifted, and its size is enlarged when an entrance slit crops the beam. This is the usual case when a slit is used to select the coherent fraction. The pairing of two focusing elements (mirrors, lenses or transfocators) to ensure a fixed focal position is also analyzed. Our results show that the image of a partially coherent source, such an undulator in a low-emittance storage ring, is a non-trivial function of the aperture used to control the coherence fraction.}

\begin{document}

\bibliographystyle{eplbib}
\maketitle

\section{Introduction}
\label{sec:introduction}
Fourth-generation storage-ring-based X-ray synchrotron sources deliver photon beams with high brilliance and coherence. Although the transverse coherence of these beams is highly improved in the horizontal direction, 
the overall coherent fraction is of the order of a few per cent for hard X-rays ($>10$~keV). Beamline optical elements such as pinholes and slits are then used to improve coherent fraction to values typically $>80 \%$ needed for applications exploiting coherence, such as X-ray photon correlation spectroscopy, coherent diffraction imaging, propagation-based phase-contrast imaging, and ptychography \cite{paganin_book}. The coherent beam interaction with the optical elements produce diffraction, thus modifying the beam characteristics and also affecting the beam focusing, as discussed here.

Let us consider the case of an ideally focusing system of focal length $f$ (made by a mirror or lens that focuses the source into the image plane). The position of the focus with respect to the focusing element $q$ is given, in the framework of geometric optics, by the lens equation $f^{-1}=p^{-1}+q^{-1}$, with $p$ the source-element distance. If the numerical aperture (NA) of the beam is reduced (for example by the finite dimension of the lens or mirror, or by using a slit) the location and dimension of the focus change as a result of diffraction. It has been shown \cite{Tanaka:85} that the focal position of a Gaussian beam moves towards the lens position when the NA decreases. This shifting of the focal position is relevant for synchrotron beamlines, as demonstrated by Westfahl {\it et al.} \cite{westfahl}. These authors noticed a shift of the horizontal focal position upstream from the position given by geometrical optics (Fig. 7 ibid.) when the horizontal acceptance is reduced by a slit. 

The diffraction effect (due to the slits and finite size of the optical elements) not only shifts the focal position (up- or downstream), but also changes the focal dimensions. These facts must be taken into account in beamlines in fourth-generation synchrotron sources. They are critical when designing beamlines with several coupled focusing elements. We studied this phenomenon in the context of the project for the new ``EBSL1'' beamline at the upgraded EBS-ESRF storage ring. This beamline will produce highly coherent beams of variable cross section at the sample position. Two refractive systems (transfocators) will be paired to allow a varying focal size, whereas a slit placed upstream from the transfocators is used to control the coherent fraction. The optical matching of the transfocators is strongly correlated to the slit aperture (or coherent fraction). The performances of such systems are calculated in the framework of the partially coherent optics using a fast algorithm for decomposition of the undulator radiation in coherent modes that are propagated along the beamline using the algorithms presented in \cite{multioptics}. The suitability of this method is discussed (ibid.) and compared with other partial coherence simulation algorithms. The key point is to treat separately the horizontal and vertical planes, therefore working with one-dimensional wavefronts.



\section{Focal position produced by a single focusing element}
\label{sec:onelens}

Let us consider an aperture of dimension $a$ placed between the source and a focusing optical element (here we consider a lens), at a distance $p_a$ from the lens ($p_a < p$). The lens is set to focus the source into an image plane at $q$ downstream from the lens. Following the geometrical optics, the position of the focal plane is defined by the lens equation (which does not depend on the aperture $a$), and the focal size is $M \times s$  (considering the aperture size $a$ is larger than the source size $s$). $M = q/p$ is the optical magnification. If $a$ is smaller than the beam size, the aperture obscures part of the source reducing the beam intensity. However, the focal size and focal position (where the beam waist is found) are not modified. When the beam has a high coherence, these results predicted by geometrical optics are insufficient \cite{hierarchical}. This section presents a numerical study of the position and size of the beam waist for different apertures taking into account the partial coherence of undulator emission. We simulated a U18 undulator (period $\lambda_u=\SI{18}{\milli\meter}$) with $N_u=138$ periods. The gap is tuned to have the first harmonic at $E=7$~keV (deflecting parameter $K=1.851$). We consider a Be lens with parabolic profile and radius at the apex $R=\SI{0.2}{\milli\meter}$ ($f=\SI{14.35}{\meter}$ at 7~keV), located at a distance $p=\SI{65}{\meter}$ from the source. A slit of variable aperture $a$ is placed at $p_a=\SI{29}{\meter}$ upstream from the lens. 

We first suppose an ideal storage ring with zero emittance, therefore producing a fully coherent emission. The beam at the slit plane has a full-width at half-maximum (FWHM) $a_\text{FWHM}=~\SI{565}{\micro\meter}$. The aperture $a$ takes different values from fully opening (the whole beam passes the slit therefore the transmittivity is $T=1$) to a very narrow aperture ($T\ll1$). The refracted beam is analyzed at different positions from the lens by calculating the FWHM of the intensity distribution. At the beam waist the FWHM presents a minimum. 
\begin{figure*}[t]
\hspace*{-1.5cm}
\centering
\includegraphics[width=1.15\textwidth]{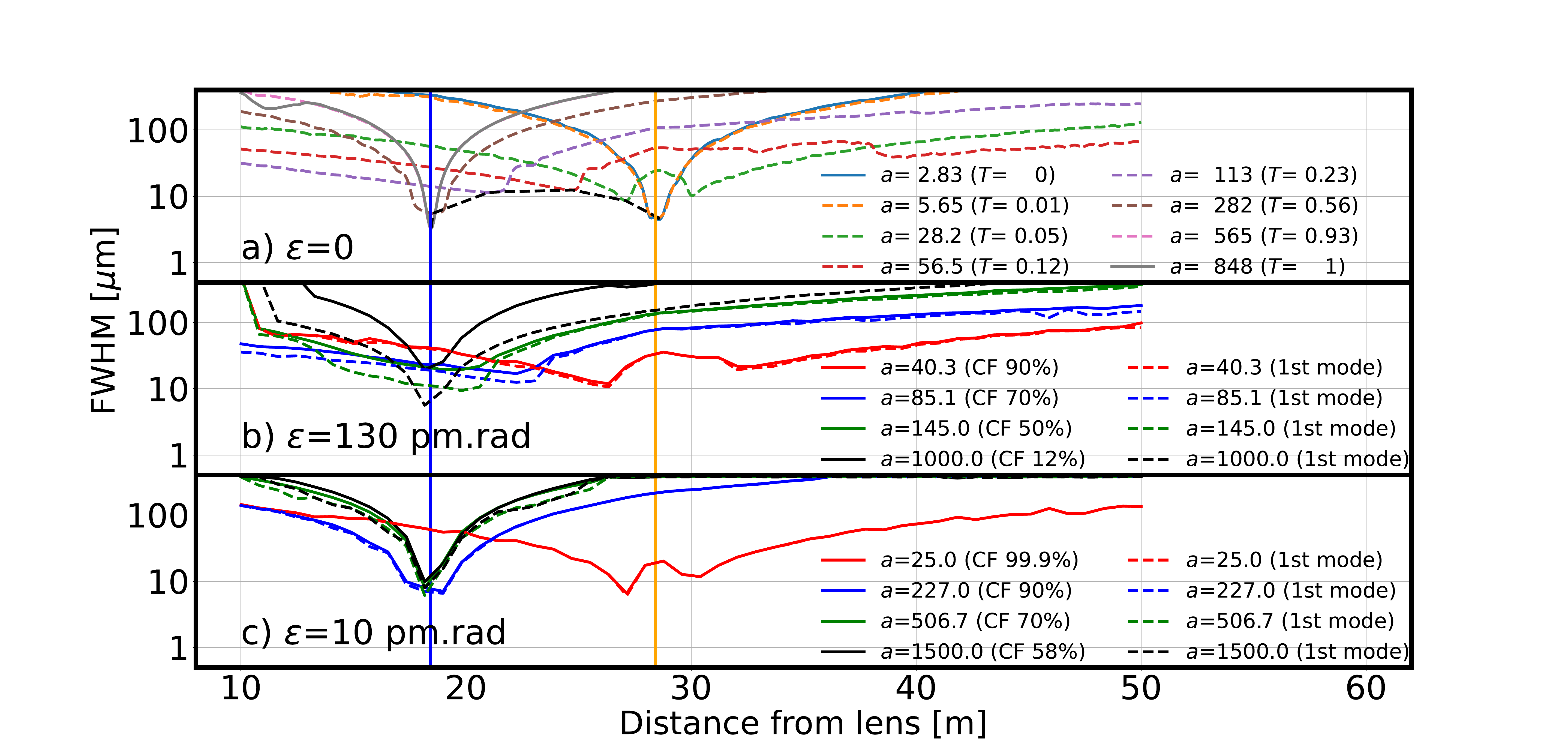}
\caption{Evolution of the size of an undulator beam cropped by a slit and focused by a lens of $R=\SI{0.2}{\milli\meter}$.
The plots show the FWHM of the beam as a function of the distance from the lens for different values of the aperture $a$ (in \SI{}{\micro\meter}). 
a) Zero emittance case ($\varepsilon=0$, full coherence),
b) horizontal emittance ($\varepsilon=130$~pm rad) and c) vertical emittance ($\varepsilon=10$~pm rad) for the EBS-ESRF storage ring.
The focal positions given by geometrical optics when the source is considered at either the undulator position (blue vertical line at $q$) or at the slit position (orange vertical line at $q_a$) are marked.
Full lines in b), c) correspond to partial coherent (multi-mode) beam, whereas the dashed lines correspond to the full coherent beam (first mode).
}
\label{fig:oneTFund}
\end{figure*}
Fig.~\ref{fig:oneTFund}a shows the evolution of the coherent beam size after being focused by the lens. It also shows markers for the focal positions predicted by the lens equation:  $q=\SI{18.417}{\meter}$ considering the source at the undulator position (blue line) and $q_a=\SI{28.408}{\meter}$ if one considers the slit as an effective source (orange line). As expected, the beam waist calculated numerically converges to these respective values when the slit is opened ($T=1$, Fresnel number $N_F=78$, grey line) and almost closed ($T\approx0$, blue line). For intermediate values of $a$, the beam waist ``moves" from one case to another. For $T=0.93$ ($N_F=35$, magenta line) the cropping by the slit is negligible so the situation does not change with respect to the open slit. For $T=0.56$ ($N_F=8.7$, brown line) the waist presents a flat depression, increasing its minimum FWHM and also the depth of focus. Smaller values of $T$ and $N_F$ shift the minimum to higher distances (e.g., $T=0.12$, $N_F=0.3$,  red line), and the minima become less pronounced. A small FWHM is found close to $q_a$ (orange marker) for $T=0.05$ ($N_F=0.09$, green line), showing a twin minimum due to the interference fringes found in the intensity distributions. Both minima converge to this position for $T=0.01$ ($N_F=0.003$, orange line), and $T=0.003\approx0$ ($N_F=0.0009$, blue line). 
The numeric values of the FWHM agree with those calculated using geometrical concepts only for the limiting cases of waist at $q$ and $q_a$. 
However, for intermediate slit apertures, where the waist position is in between $q$ and $q_a$, the waist size is different to that predicted by geometric optics, and significantly higher than the limiting values at $q$ and $q_a$, showing the envelope in Fig.~\ref{fig:oneTFund}a (dashed black line).
This is an important fact: good focusing, i.e., a very small beam size is only obtained in the limiting cases of open slit and almost-closed slit. This means that for a fully coherent beam, the slit worsens the focusing. The diffraction at the slit creates a spurious divergence that affects the lens focusing. 
When the focal distance of the lens is reduced (using lenses with smaller radius, or piling several lenses), the $q$ and $q_a$ positions shift to shorter distances, and become closer one to another ($|q-q_a|$ also reduces). They converge to a single position: the lens focal length $q=q_a=f$. This happens when both source-lens and slit-lens distances can be considered infinite.

We used the EBS-ESRF emittance values\footnote{Throughout this work we used the electron beam sizes and divergences at the center of the straight section: $\sigma_x=\SI{29.7}{\micro\meter}$,
$\sigma_{x'}=\SI{4.37}{\micro\radian}$,
$\sigma_y=\SI{5.29}{\micro\meter}$,
$\sigma_{y'}=\SI{1.89}{\micro\radian}$, corresponding to beam emittances:  $\varepsilon_x=\SI{130}{\pico\meter \radian}$,
$\varepsilon_y=\SI{10}{\pico\meter \radian}$, and beta functions
$\beta_x=\SI{6.8}{\meter}$,
$\beta_y=\SI{2.8}{\meter}$.
}
to perform the coherent mode decomposition of the undulator source. The details for that are presented in \cite{multioptics}.
This is done for the horizontal and vertical directions, yielding coherent fractions (at 7 keV) $CF_h=13\%$ and $CF_v=58\%$, respectively. These values are much higher than for the old ESRF-1 source, but still too low to apply the full-coherence approximation. Therefore, we propagated a number of modes large enough to contain more than 99\% of the source intensity (36 modes in H and 8 in V). The illumination at the entrance slit plane is \SI{610}{\micro\meter} (H) $\times$ \SI{566}{\micro\meter} (V). The slit aperture is effectively used to tune the coherence of the beam: closing the slit increases the $CF$. In the limit (zero aperture) the beam after the slit is fully coherent ($CF=1$), but obviously with zero transmittivity ($T=0$). The choice of the right slit aperture comes from a compromise between coherence and flux.
Figures~\ref{fig:oneTFund}b and~\ref{fig:oneTFund}c show the focal size along the optical axis for several apertures in the horizontal and vertical planes. We calculated the case of the partially coherent beam (multi-mode, solid lines) and also the case of a fully coherent beam (only the first coherent mode, dashed lines). We observe that the focal positions (the minima of the plot lines) do not change significantly when passing from full to partial coherence. However, the focal dimension changes significantly in the horizontal direction for cases with $CF_h\le~70\%$. In the vertical direction, where the beam at the source was more coherent, there is not much difference in sizes when going from full (dashed lines) to partial coherence (solid lines). Looking at the focal position shift versus $q_a$ (orange marker) we find important differences in the horizontal and vertical directions.
Closing the slit produces a gradual shift of the focal position from the position of the geometrical image of the undulator source (blue vertical line) to the geometrical image of the slit (orange vertical line). However, in the vertical plane, the slit crops the beam only slightly, and closing the slit to go from the source $CF_v=58\%$ to values up to 90\% does not produce any focal shift from the position of the geometrical image of the undulator source. Only when the slit is very narrow (e.g., $a_v=\SI{25}{\micro\meter}$, red curve) the focal position shifts to the locus of the geometrical image of the slit. This case is in principle not interesting experimentally as it reduces the intensity from an already quite coherent beam ($CF_v=90\%$).
In summary, for practical values of aperture selected to increase the beam $CF$ to values up to ~90\% the behaviour in vertical (V) and horizontal (H) directions is different: the better source coherence in V permits working with quite open slits thus the focusing system ``sees" the source at the undulator position, whereas in H one needs a significant crop of the beam thus shifting and enlarging the beam waist. Notice that the good coherence of the beams emitted by the EBS-ESRF and other 4$^{\text{th}}$ generation storage permits performing coherence experiments with a ``conservative" use of slit (open in V and partially closed in H). By contrast, these experiments at 3$^{\text{rd}}$ generation sources require a drastic closing of the slits to a pinhole size.


\section{Focal lengths and sizes for paired focusing elements}

Consider an ideal optical system composed of two focusing elements which have focal lengths $f_1$ and $f_2$, and which are separated by a distance $D$. Following \cite{Goodman85}, the relationship between object-to-element-1 distance $p_1$ and the element-2-to-image distance $q_2$ given by the geometrical optics is:
\begin{equation}
\label{eq:twolens}
    D-(f_1+f_2)=\frac{f_1^2}{p_1-f_1} + \frac{f_2^2}{q_2-f_2},
\end{equation}
which corresponds to an hyperbola\footnote{Expanding eq.~(\ref{eq:twolens}) one gets $f_1(D q_2 + p_1 q_2) + f_2 (D p_1 + p_1 q_2) - f_1 f_2 (D + p_1 + q_2) = D p_1 q_2$,  with horizontal asymptote at $f_{2}=((p_1+D)^{-1}+q_2^{-1})^{-1}$ and vertical asymptote at $f_{1}=(p_1^{-1}+(D+q_2)^{-1})^{-1}$.

} in the ($f_1,f_2$) plane.
The global magnification is the product of the magnification of the individual focusing elements.
\begin{equation}
\label{eq:magnification}
    M=M_1 M_2=\frac{1-q_2/f_2}{1-p_1/f_1}.
\end{equation}

\begin{figure}[htbp]
~~~~a)~~~~~~~~~~~~~~~~~~~~~~~~~~~~~~~~~~~~~~~~~b) \\
\hspace{-1.1cm}
\includegraphics[width=0.24\textwidth]{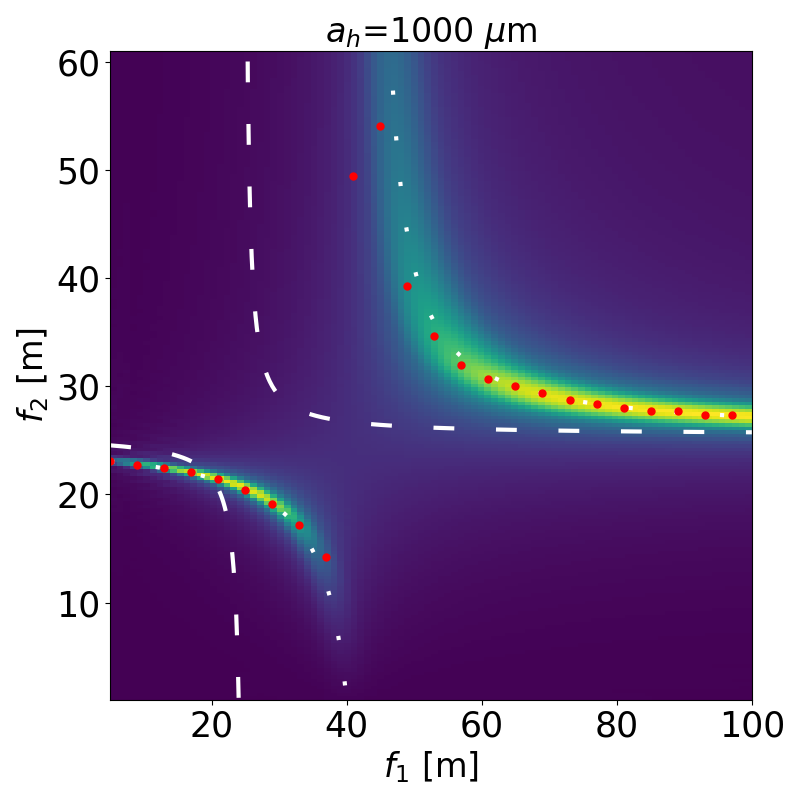}
\includegraphics[width=0.24\textwidth]{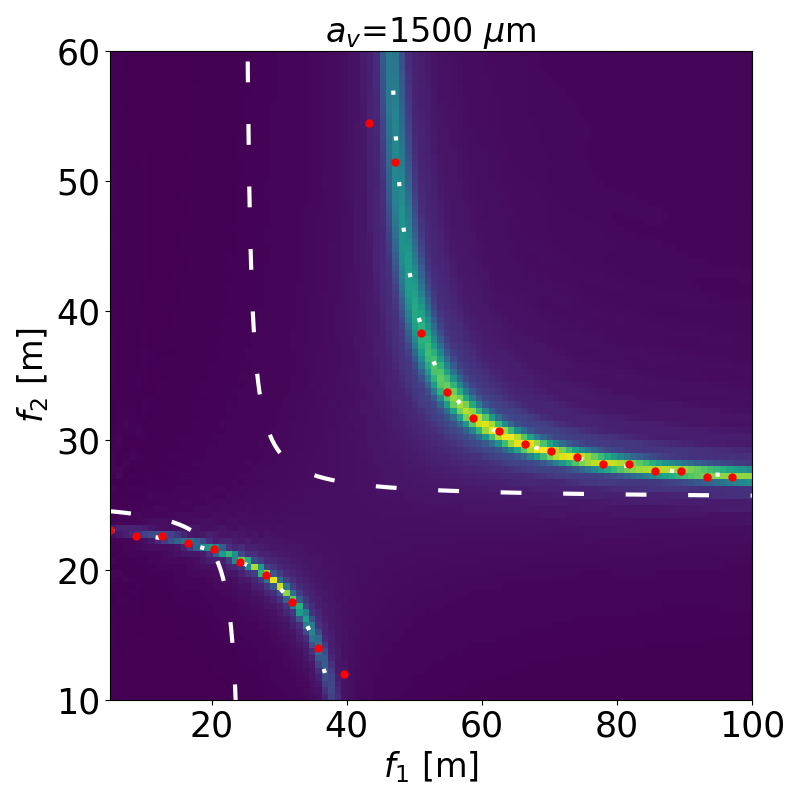}\\
\includegraphics[width=0.24\textwidth]{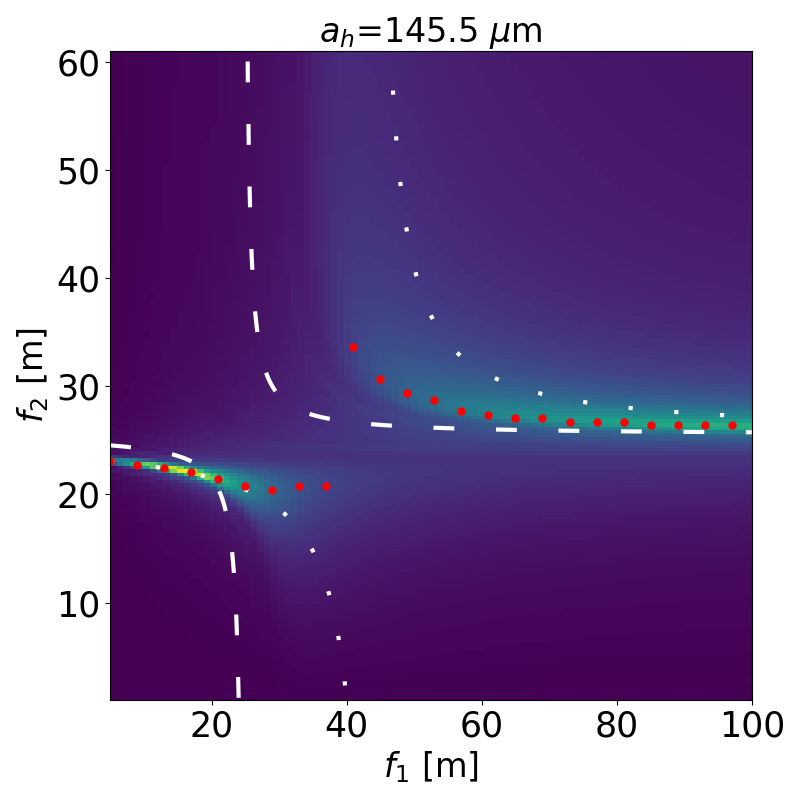}
\includegraphics[width=0.24\textwidth]{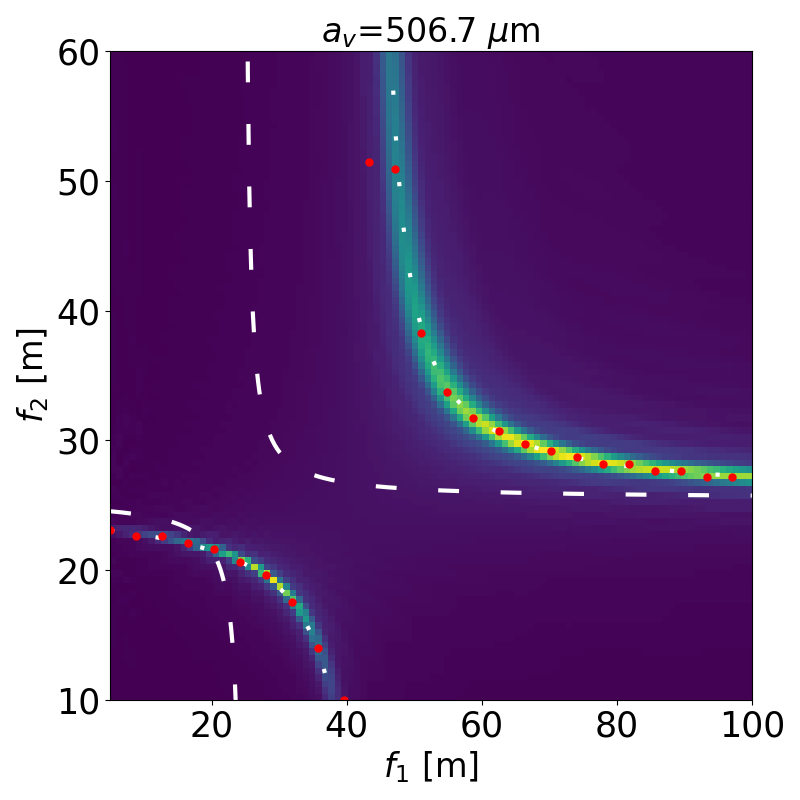}\\
\includegraphics[width=0.24\textwidth]{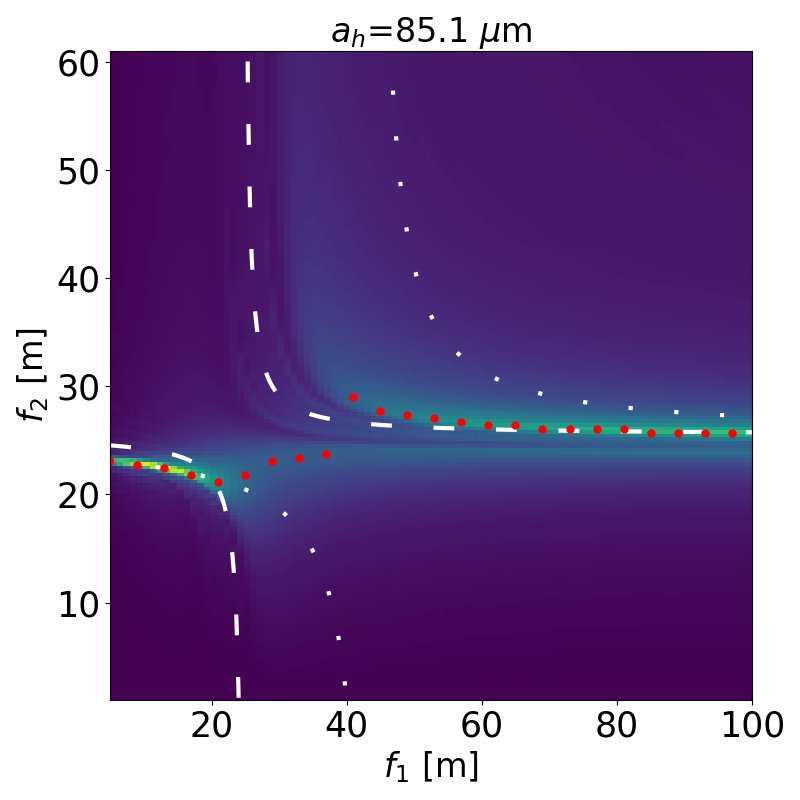}
\includegraphics[width=0.24\textwidth]{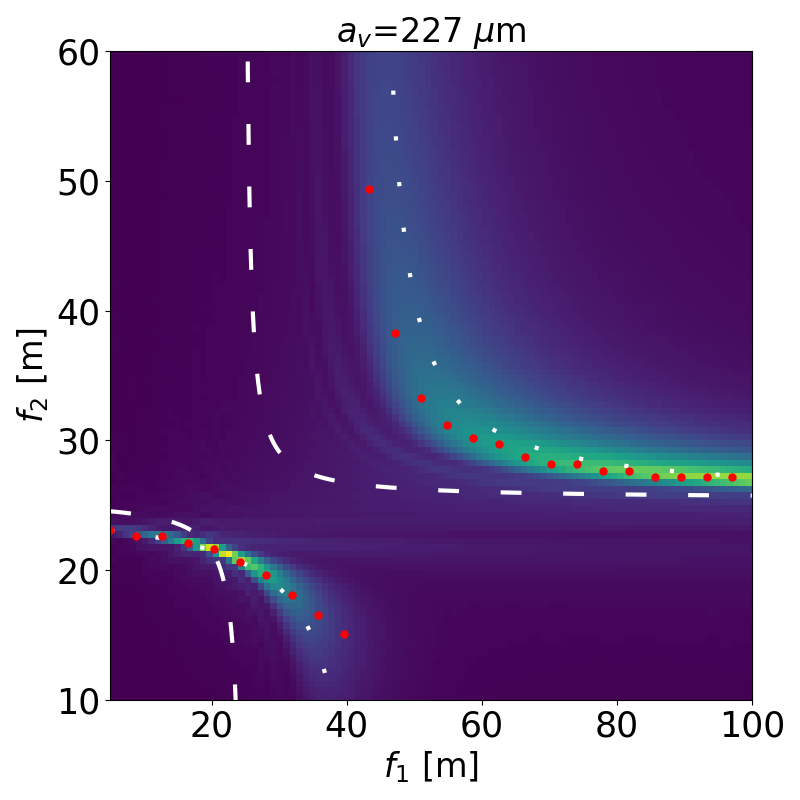}\\
\includegraphics[width=0.24\textwidth]{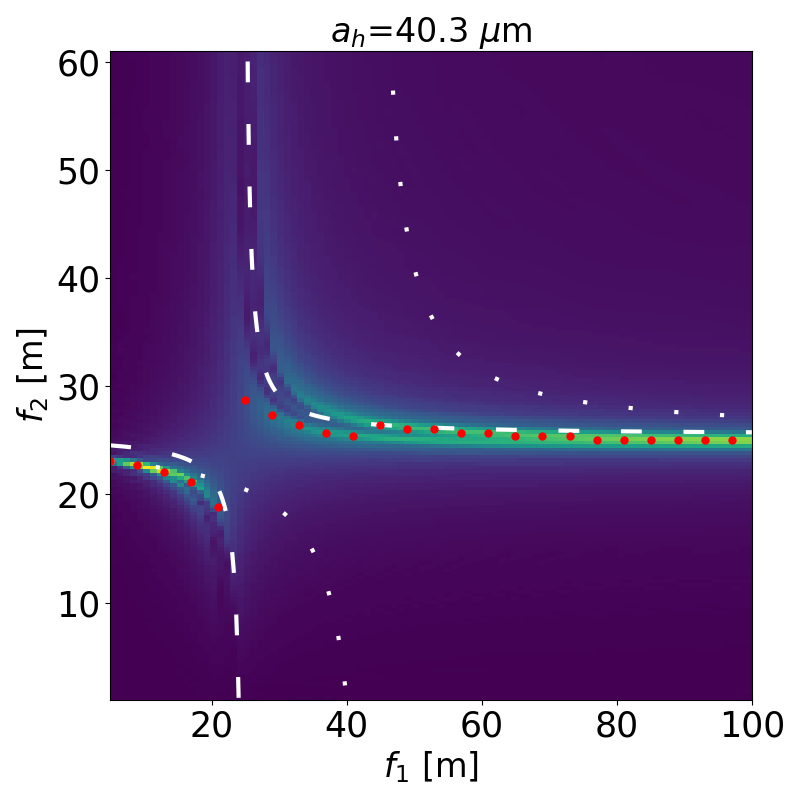}
\includegraphics[width=0.24\textwidth]{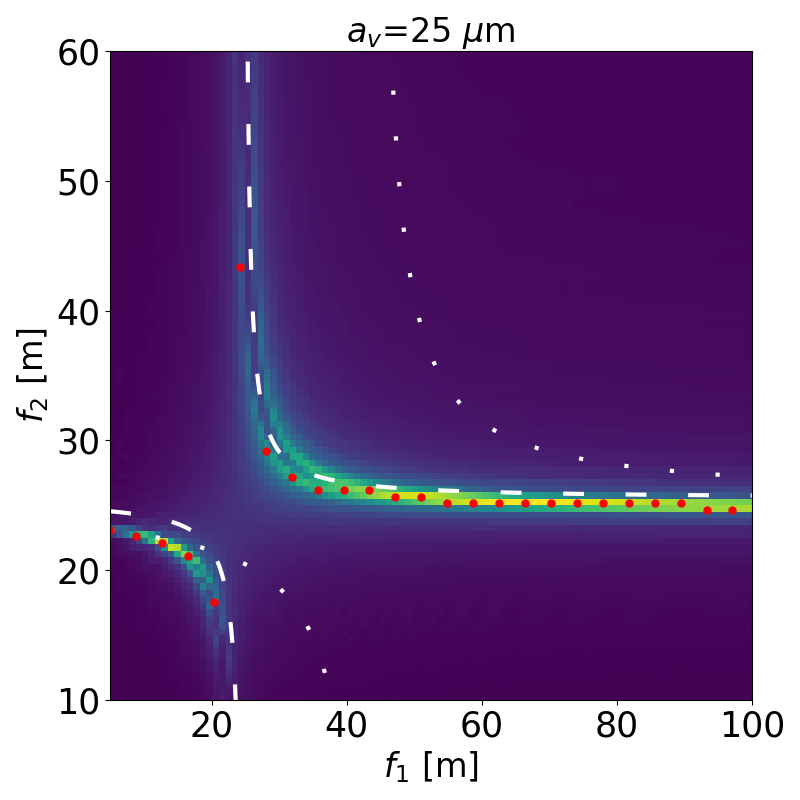}
\caption{
        \label{fig:f1f2map}
    Maps of beam peak intensity in the horizontal plane (column a) and vertical plane (b) at the sample position obtained by two lenses of variable focal lengths ($f_1$,$f_2$).
    Slit values (in \SI{}{\micro\meter}) are $a_h=[1000, 145.5, 85.1, 40.3]$ (a) and $a_v=[1500, 506.7, 227.0, 25]$ (b), corresponding to transmitted coherence fraction
    $CF_h=[0.13, 0.5, 0.7, 0.9]$ and
    $CF_v=[0.58, 0.7, 0.9, 0.999]$, respectively.
    The analytical
    hyperbolic trajectories from geometric optics (eq.~(\ref{eq:twolens})) are over-plotted in white (dotted if the source is at the undulator, and dashed if the source is at the slit). We represented in red dots some optimum pairs ($f_1,f_2$) that guarantee that the beam waist is at the focal plane.
    }
\end{figure}
\begin{figure}[H]
a)\\
\hspace{-3cm}
    \includegraphics[width=0.49\textwidth]{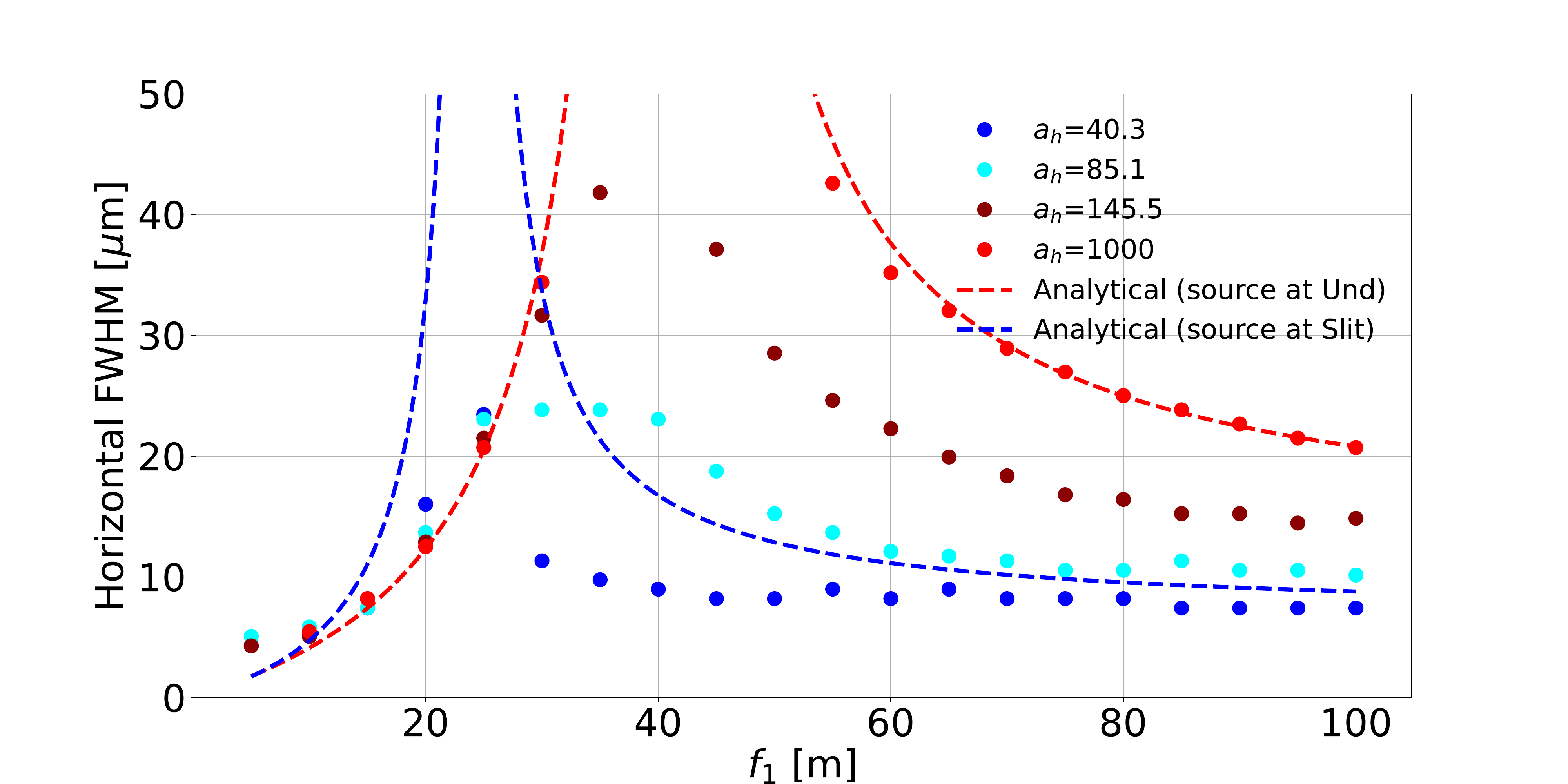}\\
b)\\
\hspace{-2.0cm}
    \includegraphics[width=0.49\textwidth]{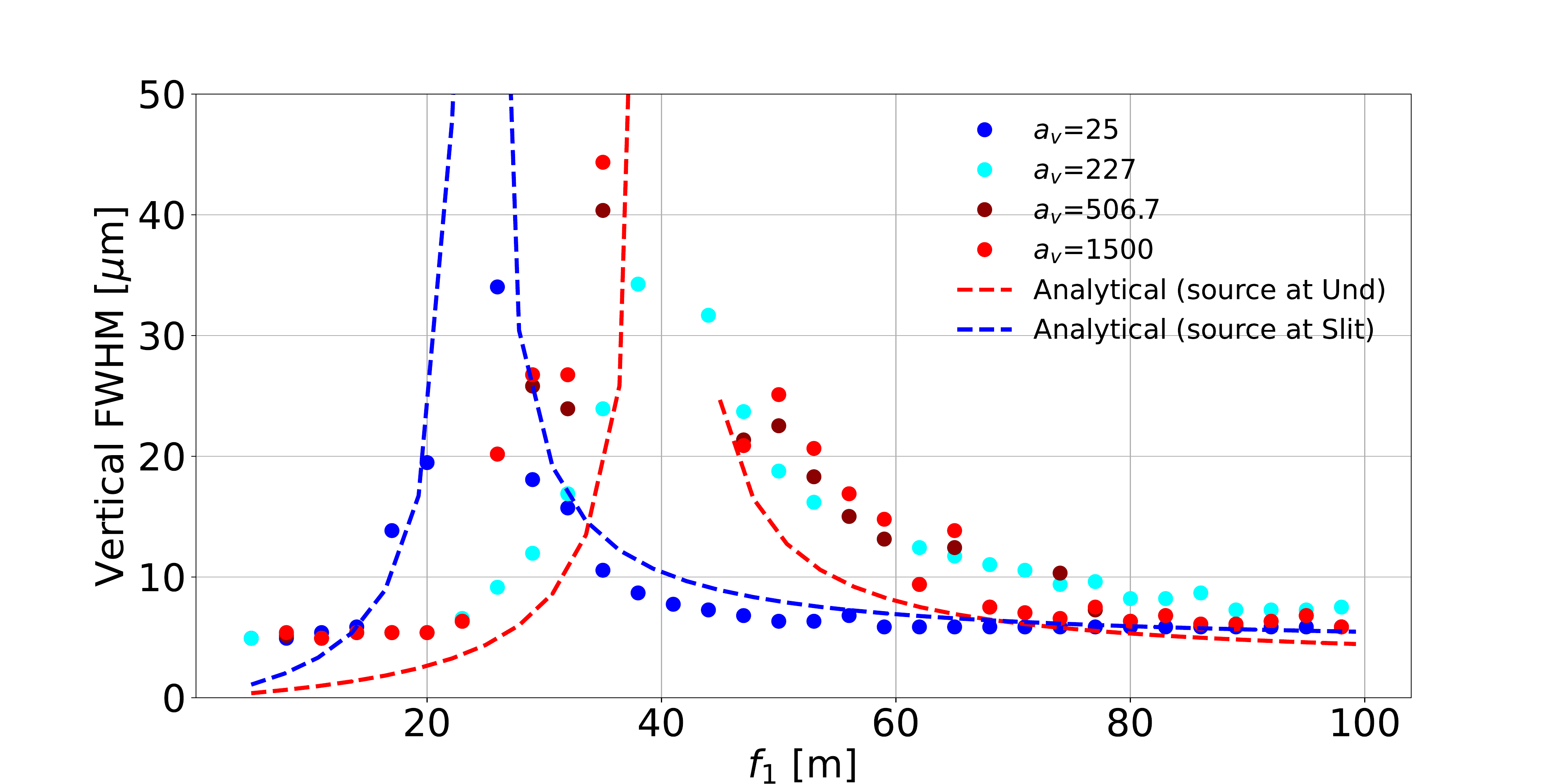}

    \caption{
    \label{fig:focalSizes}
    Size of the image in a) horizontal and b) vertical planes for different slit apertures.
    }
\end{figure}

The magnification is not dependent on $D$, however, the optical throw (length of the optical system $L=p_1+D+q_2$) will change if one changes the focal distances of the elements. For a constant $L$ one can change magnification by changing the inter-element distance $D$ (zoom effect). In a synchrotron beamline, the transfocators are refractive focusing elements that allow changing the magnification by varying their focal lengths (by adding or removing lenses) \cite{Vaughan:kv5084}. Alternatively, two Kirkpatrick-Baez systems with bendable mirrors or multilayers can be used. 

We consider an optical system made by two paired transfocators to get a variable magnification (variable focal size) in a fixed location (the sample position). The configuration match the requirements of the EBS-ESRF EBSL1 beamline. 
The source is a U18 undulator tuned at a photon energy of \SI{7}{keV} (as used before). A slit of aperture $a_h \times a_v$ is placed at \SI{36}{\meter} from the undulator.  
We represent the two transfocators by two real beryllium lenses of variable curvature radius $R_1$ and $R_2$ that correspond to focal distances $f_1$ and $f_2$, with $f_{1,2}=R_{1,2}/(2 \delta)$, $\delta=6.96\times10^{-6}$ for Be at \SI{7}{keV}. Lens-1 is placed at $p_1=\SI{65}{\meter}$ from the source, and lens-2 at $D=\SI{105}{\meter}$ downstream from lens-1. The image plane, where the sample is placed, is at $q_2=\SI{30}{\meter}$ downstream from the lens-2. The optical throw is $L=p_1+D+q_2=\SI{200}{\meter}$. The slit is $p_a=\SI{29}{\meter}$ upstream from lens-1 (i.e. at \SI{36}{\meter} from the undulator). 

For a given value of $f_1$ and the the fixed distances ($p_1$, $D$ and $q_2$), $f_2$ can be calculated analytically with eq.~(\ref{eq:twolens}). The trajectory pairs ($f_1,f_2$) obtained in this way guarantee that the focus is at the sample position ($L=\SI{200}{\meter}$ from source). However, as shown in the previous section, these results of the geometrical optics are not exact when partially coherent beams are cropped by a slit. 
Figure~\ref{fig:f1f2map} shows the results of the numerical search of the $f_2$ to produce the waist at the sample plane. It is obtained by performing a map of the on-axis intensity $\mathcal{I}_0$ of the beam at the sample plane versus $(f_1,f_2)$. Each pixel corresponds to a simulation to compute the pattern at the image plane, considering partial coherence, i.e., propagating many coherent modes. For a given $f_1$ value, one can select the $f_2$ value that produces the best focus at the sample position (picking the maximum of $\mathcal{I}_0(f_2)$, which also corresponds to the minimum FWHM). 
The $\mathcal{I}_0(f_1,f_2)$ maps depend strongly on the slit aperture. In the horizontal direction, the low $CF_h$ of the source (13\%) is strongly increased when the slit is being closed, producing an appreciable displacement of the pattern in Fig.~\ref{fig:f1f2map}a, thus requiring a fine tuning of $f_2$ to keep the beam focused at the sample plane. In the vertical direction, the source is more coherent ($CF_v=58\%$), and closing the slit does not 
shift the waist position until reaching $CF_v\approx90\%$. However, when the slit is almost closed ($a_v=\SI{25}{\micro\meter}$) acting as a pinhole the patterns approach the ($f_1,f_2$) trajectories resulting from the geometrical optics considering the source at the slit position.

Figure~\ref{fig:focalSizes} shows the calculated focal sizes. Analytical values (dashed lines), based on geometrical optics, i.e. eq.~(\ref{eq:magnification}), are close to numeric values of beam sizes for the limiting cases of open slit (source at the undulator) or almost-closed slit (source at the slit). For other useful cases where the slit crops partially the beam, the numerically calculated values should be used. Here, the focal sizes cannot be predicted by geometrical optics (also seen in Fig.~\ref{fig:oneTFund}a), and cannot be simply interpolated from the sizes corresponding to the limiting slit openings. The size values calculated analytically have very limited applicability for practical cases where the beam is partially cropped by the slit. 

In the horizontal direction, for slit apertures $a_h>\SI{40}{\micro\meter}$ the sizes change considerably for small changes in $a_h$. The focusing characteristics of the system highly depend on the diffraction effects produced at the slit. For the slit $a_h=\SI{40.3}{\micro\meter}$ ($CF_h=90\%$) the focal size can vary from roughly 10 to 50 microns. In the vertical direction the focal size can be changed from 5 to 100~\SI{}{\micro\meter} at $CF_v=90\%$.

In summary, it is shown that a slit that crops a partially coherent X-ray beam induces a diffraction effect in the beam originating an appreciable change in the focal characteristics (position and dimensions). In the case that two (or more) focusing elements are used, the basic concepts from geometric optics are insufficient to predict the conditions to pair them (define the focal lengths) for creating a focus in a precise position. Numeric methods for partially coherent optics based on coherent mode decomposition and wavefront propagation (as described in \cite{multioptics}) are used to predict the focal lengths of the focusing elements and the resulting image size.   
The data that support the finding of this study are openly available\footnote{https://github.com/srio/paper-transfocators-resources}.








\bibliography{sample}

\begin{thebibliography}{1}
\expandafter\ifx\csname url\endcsname\relax\def\url#1{\texttt{#1}}\fi

\bibitem{paganin_book}
\Name{Paganin D.~M.} \Book{Coherent X-Ray Optics} (Oxford University Press)
  2006.

\bibitem{Tanaka:85}
\Name{Tanaka K., Saga N. \and Hauchi K.} \REVIEW{Appl. Opt.}{24}{1985}{1098}.
\newline\url{http://ao.osa.org/abstract.cfm?URI=ao-24-8-1098}

\bibitem{westfahl}
\Name{Westfahl, Jr H., Lordano~Luiz S.~A., Meyer B.~C. \and Meneau F.}
  \REVIEW{Journal of Synchrotron Radiation}{24}{2017}{566}.
\newline\url{https://doi.org/10.1107/S1600577517003058}

\bibitem{multioptics}
\Name{Sanchez~del Rio M., Celestre R., Reyes-Herrera J., Brumund P. \and
  Cammarata M.} \REVIEW{Journal of Synchrotron Radiation}{29}{2022}{1354}.
\newline\url{https://doi.org/10.1107/S1600577522008736}

\bibitem{hierarchical}
\Name{Sanchez~del Rio M., Celestre R., Glass M., Pirro G., Herrera J.~R.,
  Barrett R., da~Silva J.~C., Cloetens P., Shi X. \and Rebuffi L.}
  \REVIEW{Journal of Synchrotron Radiation}{26}{2019}{1887}.
\newline\url{https://doi.org/10.1107/S160057751901213X}

\bibitem{Goodman85}
\Name{Goodman D.~S.} \REVIEW{Appl. Opt.}{24}{1985}{1732}.
\newline\url{http://ao.osa.org/abstract.cfm?URI=ao-24-12-1732}

\bibitem{Vaughan:kv5084}
\Name{Vaughan G. B.~M., Wright J.~P., Bytchkov A., Rossat M., Gleyzolle H.,
  Snigireva I. \and Snigirev A.} \REVIEW{Journal of Synchrotron
  Radiation}{18}{2011}{125}.
\newline\url{https://doi.org/10.1107/S0909049510044365}

\end{thebibliography}

\end{document}